\documentclass[aps,prb,superscriptaddress,twocolumn,showpacs]{revtex4}
\usepackage{graphicx}
\usepackage{wasysym}

\newcommand{\be}{\begin{equation}}
\newcommand{\ee}{\end{equation}}
\newcommand{\bea}{\begin{eqnarray}}
\newcommand{\eea}{\end{eqnarray}} 
  
\newcommand{\fra}[2]{\hbox{${#1\over #2}$}} 
\newcommand{\les}{\ell_{\hbox{\tiny ES}}}
\newcommand{\ld}{\ell_{\hbox{\tiny D}}}       
\newcommand{\ttr}{\tau_{tr}}
\newcommand{\tdep}{\tau_{dep}}
\newcommand{\tres}{\tau_{res}}

\newcommand{\Ndis}{N_{\hbox{\tiny dis}}}    
\newcommand{\Nrun}{N_{\hbox{\tiny run}}}    
\newcommand{\Nis}{N_{\hbox{\scriptsize is}}}    
\newcommand{\nbb}{n_{\hbox{\tiny BB}}}    
\newcommand{\Nbb}{N_{\hbox{\tiny dis}}^{\hbox{\tiny BB}}}    
    
\newcommand{\kb}{k_{\hbox{\tiny B}}}    
\newcommand{\pco}{p_{\hbox{\tiny CO}}}    
\newcommand{\dee}{\Delta E_{\hbox{\tiny eff}}}

\begin{document}

\title{Irreversible nucleation in molecular beam epitaxy: From theory to experiments}

\author{Paolo Politi}
\email{politi@ifac.cnr.it}
\affiliation{Istituto di Fisica Applicata ``Nello Carrara",
Consiglio Nazionale delle Ricerche, Via Panciatichi 56/30,
50127 Firenze, Italy}
\affiliation{Istituto Nazionale per la Fisica della Materia,
Unit\`a di Firenze, Via G. Sansone 1, 50019 Sesto Fiorentino, Italy}  

\author{Claudio Castellano}
\email{castella@pil.phys.uniroma1.it}
\affiliation{Dipartimento di Fisica, Universit\`a di 
Roma ``La Sapienza'', and Istituto Nazionale per la Fisica della Materia,
Unit\`a di Roma 1, P.le A. Moro 2, I-00185 Roma, Italy}

\date{\today}

\begin{abstract}
Recently, the nucleation rate on top of a terrace during the
irreversible growth of a crystal surface by MBE has been determined exactly.
In this paper we go beyond the standard model usually 
employed to study the nucleation process, and we 
analyze the qualitative and quantitative consequences of two important
additional physical ingredients:
the nonuniformity of the Ehrlich-Schwoebel barrier at the step-edge,
because of the existence of kinks, and the steering effects, due to
the interaction between the atoms of the flux and the substrate.
We apply our results to typical experiments of second layer nucleation.
\end{abstract}



\maketitle

\section{Introduction}
\label{sec:intro}

Nucleation is a key process during the growth of a crystal by
Molecular Beam Epitaxy (MBE), when the substrate is oriented along
a high symmetry direction. In that case, freshly deposited atoms
diffuse on the surface until they meet other adatoms (nucleation
process) or the step of a growing terrace (aggregation process).
Layers are completed through island coalescence and the filling 
of vacancies. 

Nucleation is irreversible if a dimer (i.e., a nucleus
formed by the meeting of two adatoms) is thermally stable.
This condition depends on the
two most important external parameters: the
temperature $T$ and the intensity $F$ of the flux.
It holds if the dissociation time of a dimer (which  
grows with decreasing $T$) is larger than the time (decreasing with $F$)
required to aggregate more adatoms to the dimer and to stabilize it.

As a matter of fact,
nucleation appears to be irreversible, in a range of temperatures
experimentally accessible, for several experimental systems as
Pt/Pt(111),\cite{Pt} Ag/Ag(100),\cite{Ag} Fe/Fe(100),\cite{Fe}
and Ag/Ag(111).\cite{2ndAg}

Except for the submonolayer regime, where adatoms diffuse on the
substrate, nucleations occur on terraces, and most of them
take place on `top terraces',\cite{PRE} which are defined as those
surrounded by a single closed descending step. 

In some recent papers~\cite{PRL,PRE} we have studied
the nucleation process on top of a terrace: 
we have evaluated the number $\omega$ of nucleation events 
per unit time (nucleation rate) and their spatial distribution.
The total nucleation rate $\omega$ is of great importance in several respects: 
it determines the typical distance between nucleation centers in the
submonolayer regime, the so-called diffusion length 
$\ld$ (see Ref.~\onlinecite{libroPV}); it allows 
to extract the additional energy barrier for interlayer diffusion
[Ehrlich-Schwoebel (ES) barrier] from an experiment of second layer
nucleation;\cite{2ndAg} it is the main factor 
determining the stable or unstable
character of growth during its first stages.\cite{libroMK} 

In Sec.~\ref{sec:nuc_rate} of this paper we report the formula
giving the exact value of the nucleation rate $\omega$ that was
derived previously.\cite{PRL,PRE} Such a formula depends on the
quantity $W$, the probability that two adatoms deposited simultaneously
actually meet.
Here we evaluate $W$ explicitly for all values of the terrace size $L$
and of the so-called ES length, $\les = \fra{\nu}{\nu'} -1$,
which measures the difference between the intra-layer hopping rate $\nu$ 
and the inter-layer hopping rate $\nu'$.
The resulting formula is extremely accurate.
In this way $\omega$ is written in full generality
in terms only of $L$, $F$, $\nu$, $\nu'$ and some geometrical constants 
that are given explicitly for the most relevant cases.  
We also provide the approximate formulas which are correct in some
limiting regimes.

Later we discuss, both from a qualitative and a quantitative point
of view, two issues which are believed to be important experimentally
and that are not included in the `standard' models used for studying
nucleation: the effects of nonuniform interlayer barriers
and of steering phenomena in the deposition flux.

The `standard' model for nucleation assumes the presence of an interlayer
barrier which is uniform along the step surrounding the top terrace.
However, even for compact terraces, kinks are unavoidably present along
the step and they are preferential sites for descending,\cite{kinks} 
because the ES
barrier is expected to be reduced by the increased number of neighbours.
The `effective' barrier $\dee$ is therefore a combination
of the barrier $\Delta E_k$ at kinks and the higher barrier
$\Delta E_0$, felt by an adatom along straight steps.
Even more importantly, since the number of kinks
is temperature and coverage dependent, $\dee$ depends on $T$ and
coverage, even if $\Delta E_k$ and $\Delta E_0$ do not.
This issue is treated in Sec.~\ref{sec:kinks}.

The `standard' model also assumes a uniform flux $F$ impinging on the
terrace. However, it has been shown\cite{steering_old,steeringCu,steeringAg}
that the attractive interaction between the incoming atom and the
atoms incorporated in the growing surface increases the
number of particles landing on a top terrace.
This may clearly affect the nucleation rate.
The relevance of this effect strongly depends
on the kinetic energy of the incoming particles, the strength of the
interaction and the angle of deposition.
A minimal model, where all these factors determine 
a single adimensional parameter $\epsilon$, is discussed in 
Sec.~\ref{sec:steering}.

In Sec.~\ref{sec:2nd} we apply our results to typical experiments
of second layer nucleation: an experiment on Pt/Pt(111) at
different CO partial pressures,\cite{2ndPt} and an 
experiment on Ag/Ag(111) at different temperatures.\cite{2ndAg}
In this way we are able to discuss the relevance of nonuniform barriers
and steering effects in two real cases.

Concluding remarks are presented in Sec.~\ref{sec:conc}.

\section{The nucleation rate}
\label{sec:nuc_rate}

In Refs.~\onlinecite{PRE} and~\onlinecite{PRL} we have evaluated
exactly the nucleation rate $\omega$ for a terrace of linear size $L$. 
This quantity is adimensional and represents 
the number of atoms along the edge, 
for a polygon-shaped island, or along the radius, for a circular island.
The total number of atoms in the island is ${\cal A}=\gamma L^2$. 
The most general expression for the nucleation rate is:
\be
\omega = {\cal F} {\tres \over \tres + \tdep} W \: ,
\label{eq:omega}
\ee
where ${\cal F}$ is the number of atoms arriving on the terrace per unit time,
$\tdep$ and $\tres$ are the deposition and residence time, respectively,
and $W$ is the probability that two adatoms, deposited simultaneously,
meet before leaving the terrace.

For a uniform flux, ${\cal F} = F{\cal A}$, 
and its inverse is the average time between deposition events, 
the deposition time, $\tdep = 1/{\cal F}$.
The residence time $\tres$ is the average time spent by a single particle
on the terrace before descending from it. 
It depends on the size $L$ of the terrace and on the strength of the additional
ES barrier, which can be quantified via the (adimensional) ES length,
$\les = \fra{\nu}{\nu'} -1$.
The rates $\nu,\nu'$ for intra and inter-layer hopping---whose precise
definition is given in App.~\ref{app:nu}---are generally 
written as a prefactor times an Arrhenius factor,
$\nu = \nu_0 \exp(-E_d/\kb T)$, $\nu' = \nu_0' \exp[-(E_d+\Delta E)/\kb T]$.
In the simple case where $\nu_0$ and $\nu_0'$ are supposed the same,
the ES length is\cite{note:-1} $\les = \exp (\Delta E/\kb T) -1$.
In terms of $L$ and $\les$, the residence time is given by:\cite{note:alpha}
\be
\tres = (\alpha\les +\beta L)L/\nu \: ,
\label{eq:tres}
\ee
which allows to distinguish in an easy way the
`weak barrier' regime ($\les\ll L$, $\tres=\beta L^2/\nu$) 
from the `strong barrier' one ($\les\gg L$, 
$\tres=\alpha L\les/\nu = \alpha L/\nu'$). 
The numerical factors $\alpha$ and $\beta$, as well as $\gamma$, 
depend on the shape of
the terrace and on the symmetry of the underlying lattice.
Their values for some relevant cases have been determined
numerically or analytically and are reported in Table~\ref{tab:par}.
In App.~\ref{app:eta} we determine analytically the geometrical
parameters relevant for a circular terrace.

It is worth noting that the parameter $\alpha$ can easily be written in
terms of $\gamma$ and other geometrical factors. 
As a matter of fact, in the limit of large ES barriers, the
residence time has two equivalent expressions. The first comes from
Eq.~(\ref{eq:tres}), $\tres=\alpha L/\nu'$.
The second is found by considering that 
the escape rate from the terrace, $\tres^{-1}$, is given by the probability
of finding an adatom on an edge site, 
times the fraction of hops leading from an edge site to the lower terrace,
times the interlayer hopping rate. The fraction of jumps leading to
the lower terrace is equal to $\Delta z/z$,
where $z$ is the coordination number of the lattice, and $\Delta z$ is
the number of missing neighbours for an edge site.
In the strong barrier regime the probability distribution of adatoms on the
terrace is uniform and hence the probability of finding an adatom on an
edge site is ${\cal P}/{\cal A}$, where ${\cal P} = \eta L$ is the
perimeter of the terrace in units of lattice sites.
Summing up, we can write
\be
\tres = {{\cal A}\over{\cal P}} {z\over\Delta z}{1\over \nu'} 
= {\gamma z\over\eta\Delta z} {L\over \nu'}
 \: , ~~~~~[\les\gg L] 
\label{eq:tres_2}
\ee
so that $\alpha = \gamma z/(\eta\Delta z)$.
The quantities $\eta$ and $\Delta z$ are reported in Table~\ref{tab:par}
(see also App.~\ref{app:eta}).

\begin{figure}
\includegraphics[angle=-90,width=8cm,clip]{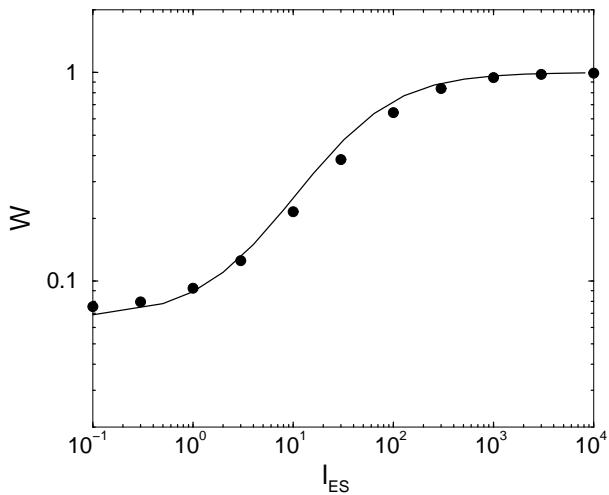}
\caption{Comparison of the exact numerical values of $W$ 
(points, see Ref.~\onlinecite{PRE}),
with the approximate analytic formula (line) given in
Eq.~(\ref{eq:W}), for a square lattice and a square 
terrace of size $L=20$.}
\label{fig:W}
\end{figure}

The quantity $W$ in Eq.~(\ref{eq:omega}), the probability that two atoms
both on the terrace actually meet,
depends on the initial spatial distributions for the two atoms,
but this dependence is very weak and can in practice be neglected.\cite{note:W}
The probability $W$ does not change much even if one atom is
taken as immobile.\cite{PRE}
In such a case, $W$ can easily be written in terms of properties
of a single random walk 
\be
W = {\Ndis(L,\les) \over {\cal A} } \:,
\ee
where $\Ndis(L,\les)$ is the number of {\em distinct} sites visited by
a single particle before leaving the terrace and clearly depends
on  $\les$ and $L$.
For extremely strong barriers, the atom is able to visit all sites,
$\Ndis\to{\cal A}$, so that $W\to 1$.
In the opposite limit of very weak barriers, it is known~\cite{Hughes} that 
$\Ndis(L,0) = N_0 L^2/\ln(L/L_0)$ (so that $W \approx 1/\ln L$).
An approximate formula interpolating between the two limits is given
in Ref.~\onlinecite{PRE}. Here we improve on that estimate by providing
an analytic expression that reproduces with very good accuracy the
value of $W$ obtained by numerical simulations {\em for all values of} $\les$.
The formula is
\be
W = { N_1\les + \displaystyle{
      {\displaystyle{N_0 L\ln(L/L_1) }\over\displaystyle{\ln(L/L_0)}}} \over
      N_1\les + \gamma L \ln(L/L_1) } \:.
\label{eq:W}
\ee
The values of $N_0, L_0, N_1, L_1$ depend on the shape of the terrace and on 
the symmetry of the underlying lattice, as $\alpha$, $\beta$ and $\gamma$ do.
Their values are given for relevant terrace shapes and lattice types in
Table~\ref{tab:par}.
\begin{table}
\caption{\label{tab:par}
Numerical values of the parameters 
$\alpha$ and $\beta$ [appearing in $\tres$, Eq.~(\protect\ref{eq:tres})],
$\gamma$ (appearing in ${\cal A}$),
$\eta$ (appearing in ${\cal P}$),
$\Delta z$ [appearing in $\tres$, Eq.~(\protect\ref{eq:tres_2})],
$N_0, L_0, N_1, L_1$ [appearing in $W$, Eq.~(\protect\ref{eq:W})].
They are given for different shapes of the terrace and for two
types of lattice: the square and the triangular one. 
They are typical of the (100) and (111) face of a cubic lattice, respectively.
For triangular ($\triangle$), square ($\Box$), and hexagonal ($\hexagon$) 
islands, $L$ is the edge of the polygon. For circular ($\bigcirc$) terrace, 
$L$ is the radius. 
The values of $\alpha$ are computed using the formula
$\alpha=\gamma z/(\eta\Delta z)$ [see Eq.~(\protect\ref{eq:tres_2})]
and agree with numerical results.The values of $N_0$, $L_0$, $N_1$ 
and $L_1$ are determined by fitting numerical results for $\Ndis(L)$ 
and $\Nbb(L)$ to their analytical expressions.}
\begin{ruledtabular}
\begin{tabular}{cccccccccc}
 & $\alpha$ & $\beta$ & $\gamma$ &  $\eta$ & $\Delta z$ &
$N_0$ & $L_0$ & $N_1$ & $L_1$ \\
\hline
(100) $\Box$ &1 &0.14 &1 & 4 & 1 &0.2 &1.0 &1.8 &0.5 \\
\phantom{(100)} $\bigcirc$ 
& $\pi/2$ & 0.5 & $\pi$ & $4\sqrt{2}$ & $\sqrt{2}$ & 0.7 & 0.5 & 3.1 & 0.3  \\
(111) $\triangle$ & 1/2 & 0.05 & 1/2 & 3 & 2 & 0.08 & 2.3 & 0.9 & 1.6 \\
\phantom{(111)} $\hexagon$ & 3/2 & 0.4 & 3 &6 & 2 & 0.7 & 0.5 & 3.0 & 0.4 \\
\phantom{(111)} $\bigcirc$ & $\pi/2$ & 0.5 & $2\pi\!/\!\sqrt{3}$ & $4\sqrt{3}$
& 2 & 0.7 &  1.0 & 3.8 & 0.2 \\
\end{tabular}
\end{ruledtabular}
\end{table}          
The derivation of Eq.~(\ref{eq:W}) is given in App.~\ref{app:Ndis} and
its accuracy is demonstrated in Fig.~\ref{fig:W}.
Notice that no parameter is fitted.

By inserting Eq.~(\ref{eq:tres}) and Eq.~(\ref{eq:W}) into
Eq.~(\ref{eq:omega}) we obtain the general fully explicit formula for the
nucleation rate. Eq.~(\ref{eq:omega}) is based on the sole hypothesis
that the deposition
time $\tdep$ is much larger than the so-called `traversal time'
$\ttr=\beta L^2/\nu$, defined as the average time an atom needs to reach the
edge of the terrace (note that $\ttr=\tres$, in the regime
$\les\ll L$).\cite{PRE}
The condition $\tdep\gg\ttr$ is equivalent\cite{PRE} to the
condition $\ld\gg 1$.
A quantitative evaluation for the experimental systems discussed 
in Sec.~\ref{sec:2nd} gives in all cases $\ttr/\tdep < 10^{-5}$.

In most cases, also the residence time is much smaller than the
deposition time. For example, for the experimental systems
studied in Sec.~\ref{sec:2nd} we have
$\tres/\tdep |_{\hbox{\footnotesize Pt}} < 5\times 10^{-3}$ and
$\tres/\tdep |_{\hbox{\footnotesize Ag}} < 10^{-2}$.
If the condition $\tres\ll\tdep$ is satisfied, the nucleation rate
is just given by the expression:
\bea
\omega &=& {\tres \over \tdep^2} \cdot W    ~~~~~ [\tres\ll\tdep]  \\
       &=& \gamma^2 {F^2\over\nu} L^5 (\alpha\les+\beta L)\cdot W \: ,
\label{eq:omega_app1}
\eea
where $W$ is given in Eq.~(\ref{eq:W}).
In the limit of strong barriers, if we approximate $W$ by one,
we have the simpler expression:\cite{KPM}
\be
\omega = \alpha\gamma^2 \; {F^2 L^5 \over \nu'} \: . ~~~~
[\tres\ll\tdep \hbox{~and~} {\nu\over\nu'}\gg L] 
\label{eq:omega_app2}
\ee

\section{The effect of non uniform barriers}
\label{sec:kinks}

An adatom leaves the terrace because it reaches an edge site
and it jumps down with a hopping rate $\nu'$:
in the usual model for nucleation, $\nu'$ is supposed to be
the same along the step edge. 
However, its value may actually be non uniform:
a growing step always has some degree of roughness,
which makes the descent preferable from certain edge sites,
and toward certain directions.
For an fcc(111) surface, 
the picture is even more complicated,\cite{libroMK} because
steps can be of two different types (A-step and B-step) and
they can both be present around the same terrace. 
In this case their barriers are different even if both types
of steps are straight.
Therefore, if the standard model is adopted, the interlayer
hopping rate should be understood as an effective one, valid at
well defined temperature and growth conditions. Any factor
able to affect the step morphology may affect $\nu'$ as well.
The `effective' character of $\nu'$ makes its separation 
in a prefactor and an exponential factor arbitrary. For this reason
we are going to assume $\nu_0'=\nu_0$.

The case of a generic distribution of ES barriers can not be treated
analytically, because it is not possible to find the general expressions
for $\tres$ and $W$. 
In the following we are assuming to be in the `strong barrier' regime,
where $W=1$ and an analytical derivation of $\tres$ is indeed possible.
In this regime, the adatom density is uniform and 
Eq.~(\ref{eq:tres_2}) is easily generalized to non-uniform barriers:
\be
\tres = { {\cal A} \over {\cal P} }{z\over \langle \Delta z\rangle}
{1\over \langle \nu'\rangle} \: .
\label{eq:tres_2_gen}
\ee
Here $\langle \Delta z \rangle$ is the number of directions leading to a hop 
downwards (averaged over edge sites), and $\langle \nu' \rangle$
is the average hopping rate.

Let us consider a simplified `kink-model':
the ES barrier is everywhere equal to the high value $\Delta E_0$,
except for special sites (and paths), where the additional barrier
takes the smaller value $\Delta E_k$.
In other words, the distribution is bimodal
\be
\Delta E = \left\{ 
\begin{array}{cll}
\Delta E_k  &  \textrm{with probability} & c_k \\
\Delta E_0  &  \textrm{with probability} & (1-c_k).
\end{array} \right.  \:
\ee

Eq.~(\ref{eq:tres_2_gen}) is valid as long as even the smallest of the barriers
is sufficiently large to ensure a uniform probability of finding the adatom
on the terrace. 
Hence the criterion for its validity is that the smallest ES length is much
larger than the linear size of the terrace,
\be
({\les})_k \equiv \textrm{e}^{-\Delta E_k/ \kb T}-1 \gg L \:.
\label{criterion}
\ee
Provided this is true, the effective barrier is determined by the relation
\be
\textrm{e}^{-\dee/\kb T} = 
c_k\; \textrm{e}^{-\Delta E_k/ \kb T} + 
(1-c_k)\; \textrm{e}^{-\Delta E_0/\kb T} \: ,
\ee
whose general solution is
\bea
&&\dee = \Delta E_k + \kb T\ln\left( {1\over c_k}\right) 
\label{eq:delta_e} \\
&&-\kb T \ln\left[ 1 + \left( {1\over c_k}-1\right) \;
\exp[-(\Delta E_0-\Delta E_k)/\kb T] \right] \: . \nonumber
\eea
In the case when the two barriers are practically the same,
Eq.~(\ref{eq:delta_e}) obviously gives
$\dee \approx \Delta E_k \approx \Delta E_0$.
In the more interesting opposite limit, when
$\Delta E_0-\Delta E_k$ is larger than $\kb T \ln (1/c_k-1)$, the 
last term on the right-hand-side is negligible, and
the effective barrier has the simplified expression,
\be
\dee = \Delta E_k + \kb T\ln\left( {1\over c_k}\right) \: .
\label{eq:delta_e_app}
\ee
Eq.~(\ref{eq:delta_e_app}) has a transparent physical meaning: adatoms
leave the terrace at kink sites only, but they feel an effective barrier
larger than $\Delta E_k$, because of the finite concentration of kinks. 
If $T=300$ K and $c_k=0.1$, the barrier increase due to this effect is 
0.06 eV.

\begin{figure}
\includegraphics[angle=-90,width=8cm,clip]{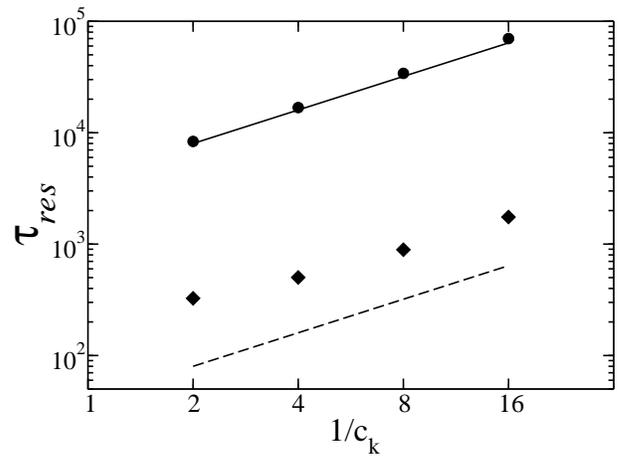}
\caption{The residence time (in units $1/\nu=1$) for an adatom 
on a square terrace of size $L=40$, as a function of the
inverse of the kink concentration (log-log scale). 
$\Delta E_0=\infty$ in all cases, while $\Delta E_k$ corresponds to 
$({\les})_k=0$  for diamonds and to $({\les})_k=99$ for circles.
Symbols refer to the exact numerical calculation of $\tres$ and
lines to the analytical approximation, 
Eqs.~(\protect\ref{eq:tres_2_gen},\protect\ref{eq:delta_e_app}).
\label{fig:tres} }
\end{figure}

What happens when the condition~(\ref{criterion}) does not hold because
$\Delta E_k$ is not large enough?
In that case the probability to find the adatom at a kink site
is suppressed and the use of Eqs.~(\ref{eq:tres_2_gen})
and~(\ref{eq:delta_e_app})
underestimates $\tres$.
This is clearly shown in Fig.~\ref{fig:tres} where we compare, for a
square terrace on a square lattice, 
numerical results (symbols) for $\tres$ 
(in units $1/\nu=1$) with the analytical approximation (lines) derived from 
Eqs.~(\ref{eq:tres_2_gen}) and~(\ref{eq:delta_e_app}).
In the case $\Delta E_0=\infty$ and $\Delta E_k=0$ the analytical
formula gives a residence time (dashed line) that is well below the
numerical results (diamonds).
In the opposite case of $\Delta E_0=\infty$ and
$({\les})_k = \exp(\Delta E_k/\kb T)-1 =99$, the condition~(\ref{criterion})
is fulfilled: numerical (circles) and analytical results (solid line) agree.

Eq.~(\ref{eq:delta_e_app}) clearly applies not only to the case of kinks,
but holds whenever there are two energy barriers for interlayer transport,
one of which is significantly larger: For example for an hexagonal terrace
on a (111) surface, surrounded by three A-steps and three B-steps. 
If $\Delta E_{A,B}$ are the ES barriers at the two different steps,
the effective barrier is equal to the smaller one plus a small
correction term,\cite{note:AB}
\be
\dee = \textrm{min}\{\Delta E_A,\Delta E_B\} +\kb T\,\ln 2 \: .
\label{eq:AB}
\ee

\section{Steering effects}
\label{sec:steering}

The flux of incoming atoms is usually supposed to be uniform.
This hypothesis is correct down to a distance of a few nanometers from
the surface, but it breaks down at smaller distances because of the
adatom-substrate interaction.\cite{steering_old}
This interaction is attractive,
so that it increases the effective flux landing on a top 
terrace.\cite{steeringCu,steeringAg}

\begin{figure}
\includegraphics[width=6cm,clip]{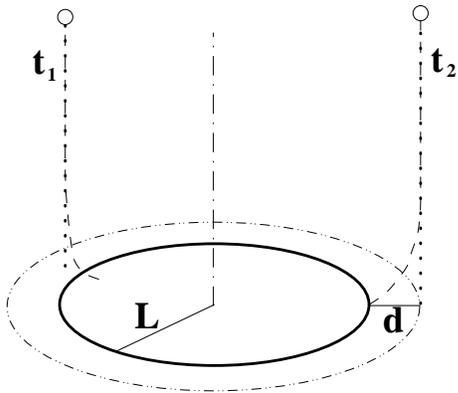}
\caption{Steering effects for normal deposition. Dotted and
dashed lines represent the trajectories without and with steering,
respectively. 
The adatoms following the trajectories $t_1$ and $t_2$
reach the terrace only because of the adatom-terrace interaction.
${\cal A}_{\hbox{\tiny eff}}$ corresponds to the area inside the 
dashed-double dotted line.
\label{fig:disegno_steering} }
\end{figure}

The effect is illustrated in Fig.~\ref{fig:disegno_steering}
for normal deposition on a circular terrace.
Because of the attractive interaction with the terrace, 
the deposited atom may deviate from its rectilinear
trajectory $t_1$ and land
on the terrace. The incoming flux ${\cal F}$, equal to $F{\cal A}=
F\gamma L^2$ in the absence of steering, is therefore increased to
the larger value ${\cal F}=F{\cal A}_{\hbox{\tiny eff}}$,
since the effective capture area of the terrace is larger.
The additional area for large $L$ is simply $d {\cal P}$, where the
quantity $d$ denotes the in-plane displacement of a particle landing
on the terrace edge (trajectory $t_2$) and ${\cal P}$ is the island
perimeter. This leads to the effective flux
\be
{\cal F} = F ({\cal A} + d\,{\cal P}) = F(\gamma L^2 + \epsilon L) \: ,
\label{eq:steering}
\ee
where $\epsilon = \eta d$ is an adimensional quantity.
The actual value of $d$ depends on the 
thermal energy of the incoming particles and on the interaction
energy between the atom and the substrate.
For oblique incidence, it also depends on the angle of incidence and it
should be meant as an average over the island perimeter.

The effect of the parameter $\epsilon$ will be illustrated in the
next Section, in the analysis of experiments on second layer nucleation.

Eq.~(\ref{eq:steering})  assumes that atoms landing on the terrace
because they are steered behave as unsteered ones. This is not exact,
because the extra flux $Fd{\cal P}$ is not deposited uniformly, but 
close to the border of the terrace. Consequently, the probability $W$ is
smaller than the value given by Eq.~(\ref{eq:W})
and its use leads to an overestimate of steering effects.
However, this is a second order effect that should be taken into
account only if a more rigorous theory for deriving the incoming flux
${\cal F}$ were employed.

\section{Second layer nucleation experiments}
\label{sec:2nd}

The nucleation rate $\omega$, as we have discussed in
Sec.~\ref{sec:nuc_rate}, is the rate of dimer formation on top
of a terrace of {\em fixed} size $L$.
Its direct experimental determination is very hard
because it would require large statistics
in single island measurements with a fixed size $L$. 

Let us instead describe how `second layer nucleation' experiments take place.
A first possibility (see Sec.~\ref{sec:Pt}) is to deposit a
fraction of a monolayer on the substrate and to analyze
the statistics of islands with a second nucleus on top.
A second possibility (see Sec.~\ref{sec:Ag})
consists in preparing an ensemble
of one monolayer high islands which is as uniform as possible,
that is with the smallest dispersion in size and 
island-island distance. Afterwards, a new fraction of a monolayer is
deposited and the statistics of islands with a second nucleus on top
of them is studied.

The relevant quantity is clearly the probability $p(\tau)$ that a 
nucleation event has occurred on top of an island during a time $\tau$.
For an ensemble of equivalent islands, this probability corresponds to the
fraction $f(\tau)$ of islands with a second nucleus:\cite{TDT} 

\be
p(\tau) = f(\tau) = 1 - \exp\left[ -\int_0^\tau \!\!dt\: \omega(L(t))\right]
\: .
\label{eq:f}
\ee
This expression clearly shows that comparison with a second layer 
nucleation experiment requires two separate pieces of information: 
the nucleation rate $\omega(L)$ and the growth law $L(t)$ of the islands.
The former has a very general validity, because it does not depend on
the details of the experiment. It does not even change if nucleation 
occurs on top of a mound:
$\omega(L)$ is a `single-island' problem and it is not affected
by the growth dynamics of the overall surface.
The latter piece of information, 
the growth law $L(t)$, is system dependent and an
exact calculation of it is generally impossible:
its determination involves, in principle, all the surface, in much the
same way as the problem of submonolayer nucleation does.

Let us now determine in an approximate way $f(\tau)$, for a uniform
array of islands of density $\Nis$ ($\Nis$ being the number of islands
per lattice site). The growth law $L(t)$  is found according
to a simple deterministic model:\cite{TDT,KPM}
each island collects the atoms falling in the capture area $1/\Nis$.
If ${\cal A}_i=\gamma L_i^2$ is the initial area of the terrace, after a time
$t$: 
\be
{\cal A}(t) = {\cal A}_i + {Ft\over\Nis} ~, ~
L(t) = \sqrt{ L_i^2 + {Ft\over\gamma\Nis} } \; .
\ee 
Hence, if islands grow from the initial size $L_i$ to the
final size $L_f$, the fraction of them with a second nucleus on top is
\be
f = 1 -\exp \left[ - {2\gamma\Nis\over F}
\int_{L_i}^{L_f} \!\!\! dL \: L\,\omega(L) \right] \: .
\label{eq:f_L}
\ee

The integral $\int dL\: L\,\omega(L)$, appearing in Eq.~(\ref{eq:f_L}),
can not be evaluated analytically for generic ES barriers:
the reason is the complex $L-$dependence of $W$ 
[see Eq.~(\ref{eq:W})]. It is easily evaluated numerically.

It is useful to define a critical size $L_c$: it is the
(final) size of the islands, corresponding to a value $f=\fra{1}{2}$.
In other words, an island grown up to the size $L_c$ has equal
chance of having or not having a second nucleus on top.
$L_c$ is determined by the implicit equation:
\be
\int_{L_i}^{L_c} \!\!\! dL \: L\,\omega(L) = {\ln 2\over 2}{F\over\gamma\Nis}
\: .
\label{eq:Lc}
\ee
We now reanalyze in detail two second layer nucleation experiments.

\subsection{Pt/Pt(111)}
\label{sec:Pt}

This experiment is described in Ref.~\onlinecite{2ndPt} and it has 
already been discussed, according to Eq.~(\ref{eq:omega_app2}), 
in Ref.~\onlinecite{KPM}. Using that expression for $\omega$,
valid for strong barriers,
the integral $\int dL\: L\,\omega(L)$ can be easily evaluated
analytically. 
Here we do not assume that barriers are strong, we use the general
expression (\ref{eq:omega}), which will be later shown to be
equivalent to (\ref{eq:omega_app1}), because of the validity of the
relation $\tres\ll\tdep$.

In the experiment, $L_i=0$ and $L_c$ has been evaluated as follows:
after deposition of a dose of platinum, the size $L_-$ of the 
smallest island {\it with} a second nucleus on top, and the size $L_+$ of
the largest island {\it without} a second nucleus on top, have been
determined experimentally for several STM topographs.
The critical size has been approximated by the mean of their
average values:
$L_c = \fra{1}{2}(\langle L_-\rangle + \langle L_+\rangle )$.\cite{Michely}

Experiments have been performed\cite{2ndPt} at different CO partial
pressures,  $\pco$.
In Table~\ref{tab:Pt} we report $\Nis$ and ${\cal P}_c$ for the
different values of $\pco$. ${\cal P}_c$ is the critical perimeter and
it is three (six) times the critical size $L_c$, 
for triangular (hexagonal) islands.
\begin{table}
\caption{\label{tab:Pt}
The critical perimeter ${\cal P}_c$ and the density $\Nis$ of islands, for
the five different concentrations of CO.}
\begin{ruledtabular}
\begin{tabular}{cccccc}
$\pco$ ($10^{-11}$mbar) & 0.5 & 10 & 47 & 95 & 190 \\
\hline
${\cal P}_c$ & 816 & 623 & 449 & 341 & 324 \\
$\Nis\times 10^5$ & 2.75 & 2.1 & 1.83 & 1.76 & 1.65 \\ 
\end{tabular}
\end{ruledtabular}
\end{table}                    
The values of the other physical parameters entering 
in Eqs.~(\ref{eq:W},\ref{eq:f_L},\ref{eq:Lc}) are:
$F=5\times 10^{-3}$ML/s, and $\nu=\nu_0\exp(-E_d/\kb T)$,
with $\nu_0=5\times 10^{12}$s$^{-1}$, $E_d=0.26$ eV,
$T=400$ K=0.0345 eV$/\kb$.
It is worth remarking\cite{2ndPt,KPM} that island coalescence has 
already started for the smallest value of $\pco$. So, in that case
the ES barrier $\Delta E$ is overestimated. 

We now have all the ingredients in order to apply Eq.~(\ref{eq:Lc})
and recover the value of the interlayer hopping rate, $\nu'$,
the only unknown quantity.
In the hypothesis of a uniform barrier along the step,
$\nu'=\nu_0'\exp[-(E_d +\Delta E)/\kb T]$. 
Two different values will be considered for the prefactor, 
$\nu_0'=\nu_0=5\times 10^{12}$s$^{-1}$ and
$\nu_0'=2.2\times 10^{12}$s$^{-1}$. The latter value has been obtained 
by Field Ion Microscopy studies.\cite{fim}
As a general rule, if $\Delta E$ is the ES barrier obtained for a
prefactor $\nu_0'=\nu_0$, for a generic value $\nu_0'$ the ES barrier
changes to a value 
\be
(\Delta E)' = \Delta E + \kb T \ln (\nu_0'/\nu_0) \: .
\label{eq:Edinu0}
\ee
For $T=400$ K, changing $\nu_0'$ by a factor two modifies
the ES barrier by 24 meV; changing $\nu_0'$ by a factor ten 
changes $\Delta E$ by 80 meV.

\begin{figure}
\includegraphics[angle=-90,width=8cm,clip]{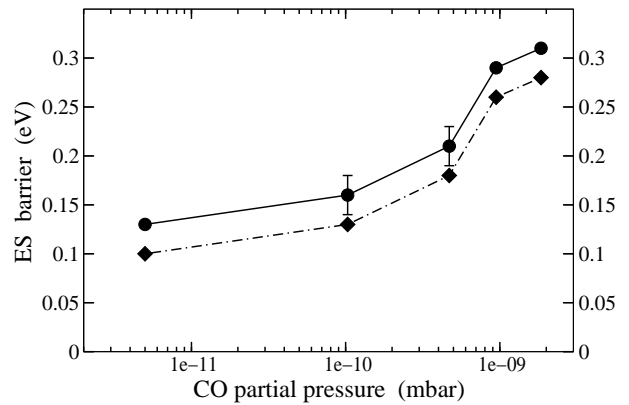}
\caption{
Values of $\Delta E$, the ES barrier, for
different $\pco$ values. Circles and diamonds refer, respectively,
to $\nu_0'=\nu_0=5\times 10^{12}$s$^{-1}$ and to
$\nu_0'=2.2\times 10^{12}$s$^{-1}$. The error bars are due to the
indetermination in the island shape.  
\label{fig:Pt_1} }
\end{figure}

\begin{figure}
\includegraphics[angle=-90,width=8cm,clip]{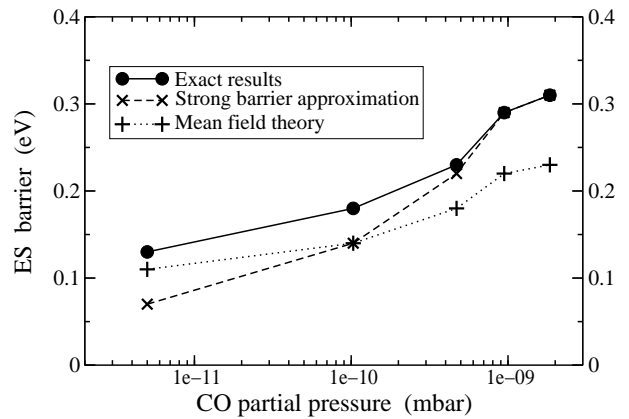}
\caption{
Results with the exact nucleation rate (for $\nu_0'=\nu_0$ and
triangular islands) are compared to Mean Field Theory and to the values
obtained from the `strong barrier' approximation for $\omega$,
see Eq.~(\protect\ref{eq:omega_app2}).
\label{fig:Pt_2}  } 
\end{figure}         
In Fig.~\ref{fig:Pt_1} we report the results for the
Ehrlich-Schwoebel barrier obtained using the exact nucleation
rate given in Eq.~(\ref{eq:omega}),
$\nu_0'$ being equal to the two values mentioned more above.
An inspection to the original STM images
(Fig.~1 of Ref.~\onlinecite{Pt}) shows that for the smallest
and the two largest values of $\pco$, islands are triangular,
whilst in the other two cases the shape is less precisely defined.
Therefore we have considered both hexagonal and triangular terraces:
the upper and lower values of the error bars correspond to such cases.
As one can see from the figure, at this temperature
the effect of the indetermination of the terrace shape is comparable
to an indetermination of $\nu_0'$ of a factor two.

In Fig.~\ref{fig:Pt_2} we compare our results
with those obtained using the strong barrier approximation,
Eq.~(\ref{eq:omega_app2}): the latter is applicable for $\Delta E\ge 0.2$ eV;
for smaller values it overestimates $\omega$ and therefore it
underestimates the ES barrier.
Mean Field Theory, also shown in the same figure, is generally incorrect.

We can now evaluate
the ratio between the residence time and the deposition time.
In the strong barrier limit, such ratio is
\be
{\tres \over \tdep} = \alpha\gamma {F\over\nu} \exp(\Delta E/T) L^3 \: .
\ee
For the largest barrier ($\Delta E=0.31$ eV), $\tres/\tdep < 0.005$.
We conclude that for analysing the experiment on platinum,
the use of formula~(\ref{eq:omega_app1}) would be legitimate.
Notice that, because of the exponential dependence of $\tres$ on $\Delta E$, 
larger values of the ES barrier quickly invalidate
Eq.~(\ref{eq:omega_app1}): $\tres \simeq\tdep$ for $\Delta E=0.4$ eV.

\begin{figure}
\includegraphics[angle=-90,width=8cm,clip]{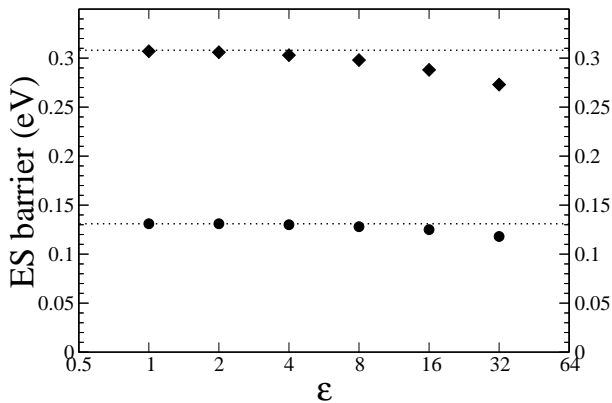}
\caption{The effect of steering on the interpretation of the second layer
nucleation experiment on platinum, Ref.~\protect\onlinecite{2ndPt}.
Note the log-linear scale.
Circles and diamonds refer, respectively, to $\pco=5\times 10^{-12}$mbar
and $\pco=1.9\times 10^{-9}$mbar.
Dotted lines mark the values of $\Delta E$ in the absence of steering.
\label{fig:dati_steering} }
\end{figure}

Let us now consider the effect of steering on the value of the ES barrier.
We repeat the evaluation of the ES barrier by considering an effective
flux ${\cal F} = F({\cal A} + \epsilon L)$ for
several values of the parameter $\epsilon$,
which measures the `strength' of steering.
Results are reported in Fig.~\ref{fig:dati_steering} (note the log-linear
scale).
For triangular terraces on an fcc(111) surface, $\eta=3$, and $\epsilon$
is three times the (maximal) in-plane
displacement of a particle landing close to the terrace edge.
It is easy to check that,
for the critical size of the terrace, the largest value of $\epsilon$
($\epsilon=32$)
corresponds to an incoming flux ${\cal F}$ increased by $60\%$
for diamonds ($\pco=1.9\times 10^{-9}$mbar) and by $24\%$ for circles
($\pco=5\times 10^{-12}$mbar).
The ES barrier is seen to decrease for strong steering, because
a larger effective flux requires a smaller barrier for producing
the same nucleation rate.
The fact that $\Delta E$ is less affected by steering when the
barrier is smaller is intuitively clear:
a smaller barrier implies a larger critical size $L_c$ and the
larger is $L$, the smaller is the effect of steering, which is an edge-effect.

\subsection{Ag/Ag(111)}
\label{sec:Ag}

The experiment on silver reported in Ref.~\onlinecite{2ndAg} is slightly 
different from the one on platinum.
First, a quantity $\theta_0=0.1$ ML is deposited.
Afterwards an equal second dose $\Delta\theta=0.1$ ML is deposited at two
different temperatures\cite{note:T} ($T=120,130$ K) 
and the fraction $f$ of covered islands
is determined as a function of the island radius $r$ (see Table~\ref{tab:Ag}).
\begin{table}
\caption{\label{tab:Ag}
Experimental data for the fraction $f$ of 
covered islands as a function of the island radius $r$, and 
theoretical results for the corresponding ES lengths, $\ell_i=\les(T_i)$.}
\begin{ruledtabular}
\begin{tabular}{ccc|ccc}
 & $T_1=120$ K & & & $T_2=130$ K & \\
 $r$(\AA) & $f$ & $\ell_1$  &  $r$(\AA) & $f$ & $\ell_2$ \\
 7.0 & 0.0  &          & 33.5 & 0.0345 & \phantom{1}5\,150 \\
11.5 & 0.02 & 300\,000 & 39.0 & 0.14 & 10\,200 \\
16.5 & 0.11 & 285\,000 & 47.0 & 0.47 & 17\,000 \\
21.5 & 0.24 & 180\,000 & 55.0 & 0.7  & 14\,800 \\
27.5 & 0.55 & 155\,000 & 63.0 & 0.85 & 11\,800 \\
32.0 & 0.88 & 200\,000 & 68.0 & 0.93 & 11\,400 \\
38.0 & 1.0  &          & 77.0 & 1.0  &       \\
\end{tabular}
\end{ruledtabular}
\end{table}                  
The density of islands is determined by the relation $\Nis=\theta_0/{\cal A}_i
=\theta_0/(\gamma L_i^2)$, and $\Delta\theta=\theta_0$ implies
$L_f=\sqrt{2}L_i$. 
Furthermore, $F=1.1\times 10^{-3}$ML/s, $\nu_0=2\times 10^{11}$s$^{-1}$ 
and $E_d=0.097$ eV.

Experimental data at different temperatures give the 
possibility---in principle---to determine both the barrier $\Delta E$
and the prefactor $\nu_0'$ of the interlayer hopping rate. 
A large debate is going on in the literature about the interpretation
of the experimental data\cite{2ndAg,KPM,Ag_debate} 
and the possibility that $\nu_0'/\nu_0\gg 1$ (see below). 
Our contribution is to analyze the data with the correct theory
and to ascertain the possible effect of nonuniform barriers and steering.

Inserting each pair of experimental values $(r,f)$ in Eq.~(\ref{eq:f_L})
yields a value for the ES length, reported in Table~\ref{tab:Ag},
and consequently for the interlayer hopping rate, $\nu'$, through the relation
$\nu'=\nu/\les$. 

For each temperature, we have used the value of $\les$ obtained from the pair
$(r,f)$ with $f$ as close as possible to $f=\fra{1}{2}$,
i.e. with $r$ as close as possible to the critical size $L_c$, because 
it is the statistically most significant point:
$f(L_f=27.5$\AA)=0.55 at $T=T_1$, and $f(L_f=47$\AA)=0.47 at 
$T=T_2$. Using the relations
\be
{\Delta E\over\kb} = 
{T_1 T_2\over \Delta T}\ln\left( {\ell_1\over\ell_2}\right) \:, \ \  
{\nu_0'\over\nu_0} = 
{1\over\ell_1} \left( {\ell_1\over\ell_2}\right)^{T_2/\Delta T} \:,
\ee
where $\Delta T=T_2-T_1$,
we finally obtain $\Delta E=0.30$ eV and $\nu_0'/\nu_0=1.9\times 10^7$.

The value found for the prefactor $\nu_0'=4\times 10^{18}$s$^{-1}$,
is huge and it is comparable to a previous analysis\cite{KPM} 
of the same data using the approximate formula (\ref{eq:omega_app2}) 
for the nucleation rate.

A possible alternative way to interpret the data is to
arbitrarily set $\nu_0'=\nu_0$, and allow the ES barrier to depend on $T$.
In this way we find $\Delta E\simeq 0.12$ eV at $T=120$ K and
$\Delta E\simeq 0.11$ eV at $T=130$ K.
Hence we find a significant difference for the barrier strength despite two
rather close temperatures.
This difference persists even if we suppose that $\nu_0'$ is 
somewhat larger than $\nu_0$ (e.g., $\nu_0'/\nu_0=10^2$).
The values of $\Delta E$ change according to Eq.~(\ref{eq:Edinu0}),
but their difference
\be
\delta (\Delta E)' = \delta (\Delta E) - \kb \Delta T
\ln(\nu_0'/\nu_0)
\label{eq:dde}
\ee
is almost unchanged, because $\kb\Delta T\approx
10^{-3}$ eV is smaller than 
$\delta(\Delta E)\equiv \Delta E$(120 K)$-\Delta E$(130 K)$\approx
10^{-2}$ eV.

We can wonder whether experimental indeterminations can explain the large
value of $\delta(\Delta E)$, or equivalently of $\nu_0'/\nu_0$.
As remarked in Ref.~\onlinecite{2ndAg}, island radii have been
determined experimentally with a precision of about 5\AA. 
Taking into account the indetermination in $r$, the average values 
of $\ell_{1,2}$ are
\be
\ell_1 \approx 180\,000 \pm 80\,000 \ \ \ \ 
\ell_2 \approx 14\,000 \pm 4\,000 \; .
\ee
However, even if we consider the smallest $\ell_1$ and the largest
$\ell_2$ [in order to minimize $\Delta E$(120 K) and to maximize
$\Delta E$(130 K)], we still have that $\delta(\Delta E)$ is of order 0.01 eV
[more precisely, $\delta(\Delta E)=8$ meV].

Let us now discuss whether the unexpected experimental results can be
the effect of non uniform barriers. 
Assuming that the prefactor $\nu_0'$ is equal or even a bit larger
than $\nu_0$, we have shown that $\delta(\Delta E)\approx 10$ meV.
If a scenario with two types of barriers applies,
Eq.~(\ref{eq:delta_e_app}) gives the relation
\be 
\delta(\Delta E) \approx \kb\bar T \ln\left(c_2\over c_1\right)
\ee
between the difference in the ES barriers and the
kink concentrations, $c_1$ and $c_2$.
In the previous equation,
$\bar T=\frac{1}{2}(T_1+T_2)$ is the average temperature. 
A value $\delta(\Delta E) \approx 0.01$ eV would therefore require
that the concentration $c_2$ of kinks at $T=130$ K is two or three times larger
than the concentration $c_1$ of kinks at $T=120$ K, which seems to be 
unlikely.\cite{note:ii}

We finally consider the effect of steering.
Increasing the value of the parameter $\epsilon$ reduces slightly
the spreading of the ES lengths (third columns in Tab.~\ref{tab:Ag}), 
but the value of $\delta(\Delta E)$ remains practically unaffected.
Alternatively, if we consider $\nu_0'$ as a free parameter, the ratio 
$\nu_0'/\nu_0$ is smaller than without steering,
but still too large:
for $\epsilon=30$ we find $\Delta E=0.23$ eV and 
$\nu_0'/\nu_0=1.5\times 10^5$.
We conclude that neither steering effects nor the presence of nonuniform
barriers are enough to allow a reasonable interpretation of the experimental
results of Ref.~\onlinecite{2ndAg}.

\section{Discussion and conclusions}
\label{sec:conc}

The present paper has three main goals.
i)~Provide a complete list of the correct formulas to be used for the
interpretation of data from second layer nucleation experiments;
ii)~Extend the `standard' model of irreversible nucleation and take
into account the effect of steering and nonuniform ES barriers;
iii)~Apply the theoretical framework to reconsider some experimental results.
Let us discuss these issues in detail.

It is now well established\cite{RM,KPM,PRL} that Mean Field 
Theory\cite{TDT} is not appropriate to study the problem of
nucleation on top of a terrace and the reason of its failure has been
clearly understood.\cite{PRE}
In Sec.~\ref{sec:nuc_rate} we give the most general formula for the
nucleation rate, Eq.~(\ref{eq:omega}), and several approximate formulas
which can be used in the relevant limits.
With respect to our recent papers\cite{PRE} on the same problem, we
are now able to provide a very good simple analytical expression for
the probability $W$ that two atoms meet (Eq.~\ref{eq:W}), valid
for any barrier strength. We also provide the numerical values for all
parameters appearing in $W$ and in $\tres$, which depend on the
shape of the terrace and on the symmetry of the underlying lattice,
see Table~\ref{tab:par} and App.~\ref{app:eta}.

Mainly because of the complicated expression for $W$, it is {\it not}
possible to integrate analytically the nucleation rate,
Eqs.~(\ref{eq:f_L},\ref{eq:Lc}), and write explicitly, e.g., the critical
size $L_c$ as a function of all the physical parameters,
$F,T,\nu,\nu',\dots$. This is not a real limitation, indeed,
because performing such one dimensional numerical integration  is 
straightforward. In the limit of strong barriers,\cite{KPM}
$\omega\sim L^5$ and the analytical integration is possible.
However, that approximation may be inaccurate, as we have shown in the case
of platinum  below $\Delta E=0.2$ eV.

In Sec.~\ref{sec:kinks} we have considered the possibility that
the additional ES barrier at steps is not homogeneous.
The case of a generic distribution of barriers can not be tackled
analytically, but if barriers are everywhere large enough to
keep the adatom density uniform, the problem is solvable.
For a simple bimodal distribution (the barrier is equal to $\Delta E_k$
in special kink sites and equal to the larger
value $\Delta E_0$ elsewhere) Eq.~(\ref{eq:delta_e_app})
indicates that the system behaves as it had a single {\it effective}
barrier $\dee$ equal to the kink barrier plus a correction depending
on the kink concentration and temperature.
This result shows that the same
experimental system displays different effective barriers at different 
temperatures, or at different growth conditions.

In Sec.~\ref{sec:steering} we have studied the effects of the
`incoming atom'-substrate interaction (steering) on the total flux landing
on a top terrace. A detailed treatment would require to consider
a realistic inter-atomic potential and to specify the energy
and the inclination of the incoming particles.
We have introduced a minimal model where the
`strength' of steering and all the above variables are included
in a single adimensional parameter, $\epsilon=\eta d$.
$\eta$ is a geometric factor
and $d$ represents the (maximal) in-plane deviation of an incoming atom.

Finally, in Sec.~\ref{sec:2nd} we reconsider the data concerning
two experiments of second layer nucleation, which had already been 
analyzed in Ref.~\onlinecite{KPM}, using the strong barrier
approximation, Eq.~(\ref{eq:omega_app2}).

In the case of Pt/Pt(111), the additional barrier varies from a
small value ($\Delta E\alt 0.1$ eV) for the clean sample, to
a value of order 0.3 eV when terrace steps are fully decorated
with CO. We stress that these values should be meant as effective ones.
If steering plays a prominent role, the effective barrier decreases,
because more adatoms are expected to land on the terrace. 
If the landing point of an adatom is displaced up to five interatomic
distances, $d=5$, the largest barrier $\Delta E=0.31$ eV 
decreases by 20 meV; if $d=10$, the reduction is twice as large.

The case of Ag/Ag(111) is very debated\cite{KPM,Ag_debate} and the 
question of the actual values of $\Delta E$ and $\nu'$ is still open,
requiring additional experiments at different temperatures.
The interpretation of the data by Bromann {\it et al.}\cite{2ndAg}
can be easily summed up.
If the prefactor $\nu_0'$ is assumed equal to $\nu_0$, the barrier
is of order 0.12 eV and it differs by about 10 meV at the two
temperatures, $T_1=120$ K and $T_2=130$ K. This difference,
$\delta(\Delta E)$, between the two barriers can be reduced if
$\nu_0'$ is allowed to increase. However, only an exponentially
large value of $\nu_0'/\nu_0$ [see Eq.~(\ref{eq:dde})]
would imply a noticeable reduction of $\delta(\Delta E)$.
If $\nu_0'/\nu_0\simeq 2\times 10^7$, the two barriers are both equal
to 0.30 eV.
Data on the fraction of covered islands at different sizes give 
different barriers. The spreading of $\Delta E$ at a given temperature
can be accounted for by the error bar on the determination of the radius $r$;
however, this is not enough to explain the difference between
$\Delta E(T_1)$ and $\Delta E(T_2)$.  
Steering effects are not able to make the difference consistent either.
In principle, we expect two different effective barriers,
at the two temperatures. Application of the formula
(\ref{eq:delta_e_app}) suggests that the value of $\delta(\Delta E)$ can
be explained by a larger kink concentration at the highest
temperature, $c_k(T_2)/c_k(T_1)\approx 2$, but it is difficult to find
a physical motivation for such a large ratio.
We conclude by remarking that a completely different 
measurement\cite{decay_Ag} of the interlayer rate in the same system,
but at a higher temperature, $T=300$ K,
gives a barrier $\Delta E=0.13$ eV and a prefactor $\nu_0'$ 
of the same order of $\nu_0$ [see also the discussion in
Ref.~\onlinecite{Ag_debate}(b)].

We believe that possible further developments in the problem of 
nucleation should include the following points: 
a direct quantitative determination
of the parameter $\epsilon$; an improvement of 
Eq.~(\ref{eq:delta_e_app}) to include the case of weak barriers;
a better assessment of the growth law $L(t)$ for the terrace;
additional quantitative and controlled second layer
nucleation experiments at different temperatures and for other systems
as well.

\appendix

\section{Intra and inter-layer hopping rates}
\label{app:nu}

The quantity $\nu$ is the total hopping rate on a flat surface.
The jump rate in a given direction is $\nu_d=\nu/z$, where $z$ is the
coordination number ($z=4,6$ for a square, triangular lattice respectively).

If an atom is on an edge site, its coordination number is reduced 
by $\Delta z$, the number of missing neighbors.
The directed hopping rate towards a missing site (i.e. towards
the lower terrace) is equal to $\nu'_d$ and it may differ from the hopping
rate, $\nu_d$, towards a site of the same terrace.
If we define $\nu'=z\nu'_d$, the absence of additional step-edge barriers
is equivalent to write $\nu'=\nu$.

Within this formalism, the escape rate from the terrace for an atom on an
edge site is equal to $\Delta z\,\nu'_d=(\Delta z/z)\nu'$.
Other authors, see e.g. Ref.~\onlinecite{libroMK}, prefer defining $\nu'$
as the escape rate ($\nu'=\Delta z\,\nu'_d$), but in that case
the absence of additional ES barriers does not match the condition $\nu'=\nu$.

\section{Calculation of some parameters for a circular terrace}
\label{app:eta}

The derivation of $\gamma$ for a circular terrace is straightforward,
because the number ${\cal A}$ of atoms contained in a circle of radius
$L$ is equal to the area $\pi L^2$ divided by the area per atom, $A_a$.
Since $A_a=1$ for a square lattice and $A_a=\sqrt{3}/2$ for a
triangular one, we have $\gamma=\pi$ in the former case,
and $\gamma=2\pi/\!\sqrt{3}$ in the latter case.
 
The derivation of $\eta$, that is, of the number ${\cal P}$ of edge sites,
is less trivial. If we change $L$ by a quantity $d_L$, the total number
of atoms changes by $d{\cal A}=2\gamma\,d_L\,L$. If we set $d_L$ so that
only edge sites are comprised in the circular ring, we have
$\eta=2\gamma\,d_L$. Let $\phi$ be the angle between the vector joining
an edge site with the center of the circle, and a nearest neighbor bond.
If $\phi=0$, $d_L=1$, whilst for a generic $\phi$, $d_L=\cos\phi$.
The value of $d_L$ entering in the relation between $\eta$ and $\gamma$ is just
the average over $\phi$ of $\cos\phi$. Because of the different lattice
symmetries, the average is evaluated between zero and $\pi/4$, for a square
lattice, and between zero and $\pi/6$, for a triangular lattice.
We therefore obtain $d_L=2\sqrt{2}/\pi$ in the former case, and
$d_L=3/\pi$ in the latter case. In conclusion, we obtain
$\eta=4\sqrt{2}$ for the (100) lattice  and $\eta=4\sqrt{3}$ for the
(111) lattice.
                                                               
The derivation of $\Delta z$, the (average) number of missing
neighbours per edge site, is easily found from ${\cal P}$.
Let us first determine the number $N_{nn}$ of nn bonds
which are cut by a circle of radius $L$. It is elementary to write
that $N_{nn}=8L$ for a square lattice and $N_{nn}=8\sqrt{3}L$
for a triangular lattice. The quantity $\Delta z$ is nothing but
the ratio between $N_{nn}$ and ${\cal P}$, so that:
$\Delta z=\sqrt{2}$ for the (100) lattice and $\Delta z=2$ for the
(111) lattice.

\section{Calculation of $\Ndis$}
\label{app:Ndis}

We want to evaluate analytically the average number of distinct 
sites $\Ndis$ visited by an adatom during its random walk on the terrace.
Let us imagine that we perform $\Nrun$ times the following procedure.
We let a random walker start from a site on the terrace and follow its
trajectory until it leaves the terrace.
For each run $r$ and each terrace site $s$ we define the quantity $n(r,s)$
to be 1 if site $s$ has not been visited during run $r$ and 0 otherwise.
The average probability that site $s$ is not visited is 
\be
n(s) = \lim_{\Nrun\to\infty} {1\over\Nrun} \sum_{r=1}^{\Nrun} n(r,s) \: .
\ee 

All runs can be grouped according to the number $F\ge 1$ of `traversals'
across the terrace, defined as follows.
Once the atom has arrived on an edge site, it has the probability
$p_{out}=\Delta z/z$ to attempt to move outside and
$p_{in}=1-\Delta z/z$ to move to another site of the terrace:
in the former case, the 
atom has the probability\cite{PRE}  $a=\les/(1+\les)$ to stay there
and the probability $1-a=1/(1+\les)$ to leave the terrace.
The first traversal starts when the atom is deposited on the
terrace and terminates when the atom reaches an edge site {\it and}
tries to leave the terrace. At this point the atom may leave the terrace,
in which case it has performed one traversal only, or it may
stay on the terrace and start a new traversal.
According to the this definition, we can write
\be
n(s) = \lim_{\Nrun\to\infty}
{1\over\Nrun} \sum_{F=1}^\infty \Nrun^F \langle n(r,s)\rangle_F \: ,
\label{eq:n_gen}
\ee
where $\Nrun^F$ is the total number of runs made up of $F$ 
traversals, and $<\cdots >_F$ is the average value
on those runs only.

The quantity $n(r,s)$ is different from zero if and only if the site $s$
has not been visited in any traversal. We can write explicitly
\be
n(r,s) = \prod_{k=1}^F n_k(r_F,s) \: ,
\label{eq:n_p1}
\ee
where $n_k$ is the variable $n$ referred to the $k-$th traversal of the run
labelled $r_F$. Inserting (\ref{eq:n_p1}) in Eq.~(\ref{eq:n_gen}) we
obtain 
\be
n(s) = \sum_{F=1}^\infty \mu(F) \langle\prod_{k=1}^F n_k(r_F,s)\rangle\: ,
\label{eq:n_gen2}
\ee
where $\mu(F)=\Nrun^F/\Nrun$ is the probability that a single run is made up
of $F$ traversals.

If a run goes on for $F$ traversals, it means that the atom has done $F$
attempts to descend, failing the first $(F-1)$ times and succeeding the last
one. Therefore,
\be
\mu(F)= a^{F-1} (1-a) \: .
\label{eq:mu}
\ee

We now make the approximation that distinct traversals are independent:
\be
\langle\prod_{k=1}^F n_k(r_F,s)\rangle \simeq
\langle n_1(r_F,s)\rangle \prod_{k=2}^F \langle n_k(r_F,s)\rangle\: .
\label{eq:app}
\ee

Using Eqs.~(\ref{eq:mu},\ref{eq:app}) in (\ref{eq:n_gen2}), we obtain
\be
n(s) \simeq \sum_{F=1}^\infty (1-a)a^{F-1} \langle n_1(r_F,s)\rangle
\prod_{k=2}^F \langle n_k(r_F,s)\rangle\: .
\ee

Let us remind the meaning of $\langle n_k(r_F,s)\rangle$. 
It is the probability that site $s$ is {\it not} visited during the
$k-$th traversal of a run composed of $F$ traversals.
This quantity does not depend on $F$. We make the approximation that
it does not depend on $k$ either, if $k\ge 2$.
In simple words, we just define $\langle n_1(s)\rangle$ for the first traversal
and $\langle\nbb(s)\rangle$ for all the subsequent traversals.
The lowerscript BB means that the traversal starts from and arrives
at the boundary of the terrace.

Within this approximation,
\bea
n(s) &\simeq & \langle n_1(s)\rangle (1-a) \sum_{F=1}^\infty 
a^{F-1} \langle\nbb(s)\rangle^{F-1} \nonumber\\
 &\simeq & \langle n_1(s)\rangle 
{ (1-a) \over 1 - a\langle\nbb(s)\rangle } \nonumber\\
&\simeq & { \langle n_1(s)\rangle \over 
1+ \les (1-\langle\nbb(s)\rangle) } \: .
\label{eq:n_fin}
\eea

The last approximation is to neglect the $s-$dependence in 
$\langle\nbb(s)\rangle$ and $\langle n_1(s)\rangle$.
Using the simple relation, $\Ndis=\sum_s (1- n(s))$, we obtain 
\be
W =  { \Ndis(L,\les) \over {\cal A} } 
  =  { \les \Nbb + \Ndis(L,0) \over 
      \les \Nbb + {\cal A} } \:
\label{eq:Wgen}
\ee
where $\Ndis(L,0)=(1-\langle n_1(s)\rangle)/{\cal A}$ and
$\Nbb(L)=(1-\langle\nbb(s)\rangle)/{\cal A}$ is the number of distinct sites
visited by a particle starting from the boundary and reaching it again.
Inserting the expression for $\Nbb(L) = N_1 L/\ln(L/L_1)$ we obtain the
explicit form of $W$
\be
W = { N_1\les + 
      {\displaystyle{N_0 L\ln(L/L_1) }\over\displaystyle{\ln(L/L_0)}} \over
      N_1\les + \gamma L \ln(L/L_1) } \:.
\ee

\begin{acknowledgments}
We gratefully thank Thomas Michely and Harald Brune for providing
the original data of the two experiments discussed in
Sec.~\ref{sec:Pt} and Sec.~\ref{sec:Ag}.
Thomas Michely and Joachim Krug are acknowledged for their critical reading
of the manuscript.
Special thanks go to Ruggero Vaia for having suggested how to determine
$\eta$ and $\Delta z$ for a circular terrace.
Part of this work was done during the International Seminar
``Models of Epitaxial Crystal Growth" held at the
Max Planck Institute for the Physics of Complex Systems, Dresden,
in March 2002.
\end{acknowledgments}


\begin{thebibliography}{99}
\bibitem{note:-1}
When the ES length is large, we can neglect the term $(-1)$ appearing in 
its general definition.
\bibitem{note:alpha}
With respect to our previous Refs.~\protect\onlinecite{PRL,PRE},
we have introduced the hopping rate $\nu$ rather than the diffusion
constant $D=\nu/4$, in the expression (\protect\ref{eq:tres}) of the
residence time. Therefore the new $\alpha$ and $\beta$ are four 
times larger than the old ones.
\bibitem{note:W}
This is not true if the particle is initially confined to the edge of the
terrace, but this case is not practically relevant.
\bibitem{note:AB}
In Eq.~(\protect\ref{eq:AB}) we are
neglecting the existence of a kink between neighbouring straight steps.
\bibitem{note:T}
Experimental data are available at $T=70$ K as well, but they can not
be used in the present context, because $f=1$ for all observed islands.
\bibitem{note:ii}
Because of kinetic effects the number of kinks
should decrease with increasing $T$. Because of thermodynamical effects,
it should increase, but $\kb\Delta T\approx 1$meV is much smaller than
the kink energy.
\end{thebibliography}
\end{document}